\documentclass[aip,reprint]{revtex4-1}
\usepackage{graphicx}

\begin{document}
%------------------------------------- Title -------------------------------------%
\title{
Simultaneous measurements of super-radiance at multiple wavelengths from helium excited states: (I) Experiment
}
%------------------------------------- Authors -------------------------------------%

\author{Kyo Nakajima}
\thanks{Present address: Institute of Materials Structure Science, KEK, Tsukuba, Ibaraki 305-0801, Japan}
\affiliation{Research Core for Extreme Quantum World, Okayama University, Okayama 700-8530, Japan}

\author{James R.\ Harries}
\affiliation{JAEA/SPring-8, Kouto 1-1-1, Sayo, Hyogo, 679-5148 Japan}	

\author{Hiroshi Iwayama}
\affiliation{Institute for Molecular Science, National Institutes of Natural Sciences, Okazaki 444-8585, Japan}
\affiliation{SOKENDAI, Nishigo-naka 38, Myodaiji, Okazaki, Aichi, 444-8585 Japan}	
        
\author{Susumu Kuma}
\thanks{Present address: Atomic, Molecular \& Optical Physics Laboratory, RIKEN, Wako, Saitama 351-0198, Japan. Corresponding author}
\affiliation{Research Core for Extreme Quantum World, Okayama University, Okayama 700-8530, Japan}
	
\author{Yuki Miyamoto}
\affiliation{Research Core for Extreme Quantum World, Okayama University, Okayama 700-8530, Japan}
        
\author{Mitsuru Nagasono}
\affiliation{RIKEN SPring-8 Center, Kouto 1-1-1, Sayo, Hyogo 679-5148, Japan}
        
\author{Chiaki Ohae}
\thanks{Present address: Department of Engineering Science, University of Electro-Communications, Chofu, Tokyo 182-8585, Japan}
\affiliation{Graduate School of Natural Science and Technology, Okayama University, Okayama 700-8530, Japan}

\author{Tadashi Togashi}
\affiliation{Japan Synchrotron Radiation Research Institute (JASRI), Kouto 1-1-1, Sayo, Hyogo 679-5198, Japan}	
        
\author{Makina Yabashi}
\affiliation{RIKEN SPring-8 Center, Kouto 1-1-1, Sayo, Hyogo 679-5148, Japan}
\affiliation{Japan Synchrotron Radiation Research Institute (JASRI), Kouto 1-1-1, Sayo, Hyogo 679-5198, Japan}	        
        
\author{Eiji Shigemasa}
\affiliation{Institute for Molecular Science, National Institutes of Natural Sciences, Okazaki 444-8585, Japan}	
\affiliation{SOKENDAI, Nishigo-naka 38, Myodaiji, Okazaki, Aichi, 444-8585 Japan}
	
\author{Noboru Sasao}
\affiliation{Research Core for Extreme Quantum World, Okayama University, Okayama 700-8530, Japan}

%------------------------------------- Affiliations -------------------------------------%

%------------------------------------- Abstract -------------------------------------%
\begin{abstract}
	In this paper, we report 
	the results of measurements of the intensities and delays of super-radiance decays from excited helium atoms at multiple wavelengths.
        The experiment was performed using extreme ultraviolet 
        radiation
        produced by the free electron laser 
        at the SPring-8 Compact SASE Source test accelerator facility as an excitation source.
        We observed super-radiant transitions on the
        $1s3p \to 1s2s$ ($\lambda=$502 nm), 
        $1s3d \to 1s2p$ ($\lambda=$668 nm), and $1s3s \to 1s2p$ ($\lambda=$728 nm) transitions.
        The pulse energy
         of each transition and its delay time were measured as a function of the target helium gas density.
        Several interesting features of the data, some of which appear to contradict with 
        the predictions of the simple
         two-level super-radiance theory, are pointed out.
\end{abstract}

%------------------------------------- Key Words -------------------------------------%

\maketitle

%------------------------------------- Section 1: Introduction -------------------------------------%
\section{Introduction}
%------------------------------------- Section 1 -------------------------------------%

Super-radiance (SR), predicted by Dicke in 1954,\cite{R.H.Dicke} belongs to a class of coherent and cooperative optical phenomena. 
One remarkable feature of super-radiance is its radiation dynamics.
In contrast to the well-known exponential decay of spontaneous emission, super-radiant decay has a pulse-like time profile, with a peak intensity proportional to $n^2$, and a delay time inversely proportional to $n$, where $n$ is the number of excited atoms involved in the relevant process.
Thus SR may be regarded as a phenomenon in which deexcitation processes are accelerated (or amplified) by the coherence developed among atoms.
Since its original prediction, many experimental and theoretical studies have been performed to investigate the detailed nature of the phenomenon.\cite{SR-review-Benedict}

%+++++++++++++++++++++ Figure 1 : helium energy level +++++++++++++++++++++%
\begin{figure}[h]
	\begin{center}
		\includegraphics[width=8.5cm]{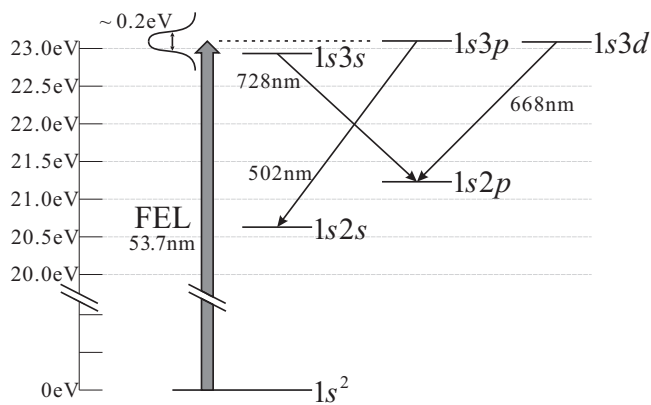}
        \end{center}
	\caption{Low-lying helium energy levels. 
	Helium atoms were excited from the ground state ($1s^2$) to the $1s3p$ excited state by 53.7 nm wavelength EUV-FEL radiation.
Three super-radiance transitions were observed, corresponding to $1s3p \to 1s2s$ (502 nm), $1s3d \to 1s2p$ (668 nm), and $1s3s \to 1s2p$ (728 nm).
The average FEL spectral profile is depicted schematically.
        }
	\label{fig:He Energy Levels}
\end{figure}
%+++++++++++++++++++++ Figure 1 +++++++++++++++++++++%

Recently, some of us reported the observation of super-radiance on the 502-nm-wavelength $1s3p \to 1s2s$ transition in helium following $1s^2 \to 1s3p$ excitation using extreme ultraviolet free electron laser (EUV-FEL) pulses from the SPring-8 Compact SASE Source test accelerator facility (SCSS).\cite{He-SR-Nagasono}
(All the states mentioned in this paper are singlet states, and we omit term symbols such as $^{1}S$ or $^{1}P$.)
Helium is the lightest rare gas atom, and its electronic excited states all have excitation energies above 20 eV. 
The ultrafast creation of a large enough population of excited helium atoms to observe super-radiance only became possible with the advent of the EUV free-electron laser.
While super-radiance was only initially observed on the $1s3p \to 1s2s$ transition, super-radiant decays in the $1s3s \to 1s2p$ and $1s3d \to 1s2p$ transitions were observed later\cite{He-SR-Harries}.
It was indicative that the phenomenon occurred in a multi-level system (see Table \ref{Table:transitions} for the parameters of related transitions). 
SR in multi-level systems, such as cascade or competitive SR, has been studied in many different experimental settings\cite{PhysRevLett.36.1035, 0022-3700-12-4-006, 0022-3700-12-7-003, Florian1984169, Kuprionis1994}, and the observed results could be broadly understood using simple two-level approximations\cite{PhysRevLett.36.1035, 0022-3700-12-4-006}.

%+++++++++++++++++++++ Table 2 +++++++++++++++++++++%
\begin{table}[h]
\begin{center}
\caption{Transitions related to the observed SR.   $\lambda$, wavelength; $A$, Einstein A-coefficient.}
\label{Table:transitions}
\begin{tabular}{c|ccc}
\hline \hline
Transition & $\lambda$ [nm] & A [$\mathrm{s}^{-1}$] & $\lambda^{2}A$ $[\mathrm{m}^2\mathrm{s}^{-1}]$ \\
\hline
$1s3p - 1s2s$ & 501.7 &	 $1.3\times10^{7}$ &		 $3.4\times10^{-6}$  \\
$1s3d - 1s2p$ & 668.0 &	 $6.4\times10^{7}$ &		 $2.8\times10^{-5}$  \\
$1s3s - 1s2p$ & 728.3 &	 $1.8\times10^{7}$ &		 $9.7\times10^{-6}$  \\
$1s3p - 1s3s$ & 7437.5 &	 $2.5\times10^{5}$ &		 $1.4\times10^{-5}$  \\
$1s3p - 1s3d$ & 95787 &	 $1.5\times10^{2}$ &		 $1.4\times10^{-6}$  \\
\hline \hline
\end{tabular}
\end{center}
\end{table}
%+++++++++++++++++++++ Table 2 +++++++++++++++++++++%

Here, we report an experimental investigation of the dependence on sample density of the competitive SR dynamics of helium atoms excited at the EUV-FEL beamline at SCSS. 
The short excitation time of the FEL ($\sim30$ fs), combined with detection on a timescale of tens of picoseconds allowed us to study the SR dynamics in detail. 
Super-radiance delay times and peak intensities for multiple-wavelength SR transitions could be extracted from the data for a range of sample densities covering almost three orders of magnitude.
The rest of the paper is organized as follows. 
In the next section, we describe the experimental apparatus and procedures.
The results and summary are given in \S 3 and \S 4, respectively.
The detailed numerical analysis of the present experiment from the view point of Maxwell-Bloch equations will be described in an accompanying paper\cite{Analysis-Paper}.

%------------------------------------- Section 2: Experiment Apparatus -------------------------------------%
\section{Experiment Apparatus}
\begin{figure}[h]
	\begin{center}
		\includegraphics[width=8.5cm]{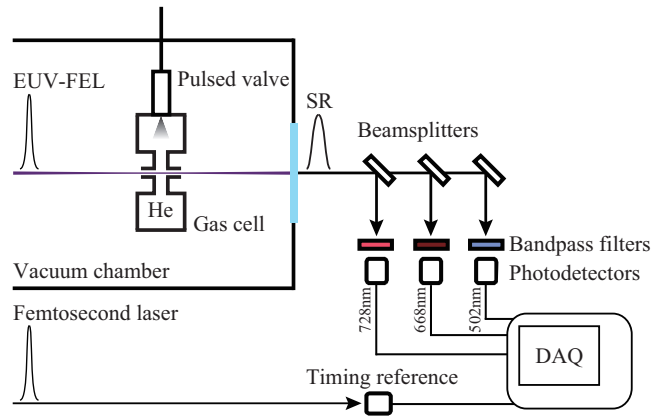}
        \end{center}
	\caption{Schematic diagram of the experimental apparatus.
        }
	\label{fig:Setup}
\end{figure}

%------------------------------------- Section 2 -------------------------------------%
The main components of the experimental apparatus are 
(1) the EUV-FEL beam line, (2) the helium gas target, (3) the radiation detectors and (4) the data acquisition system (Fig.\ \ref{fig:Setup}).
We describe below these components in turn.

%+++++++++++++++++++++ Figures 2&3 : Setup +++++++++++++++++++++%
\begin{figure}[h]
	\begin{center}
    		\includegraphics[width=8.5cm]{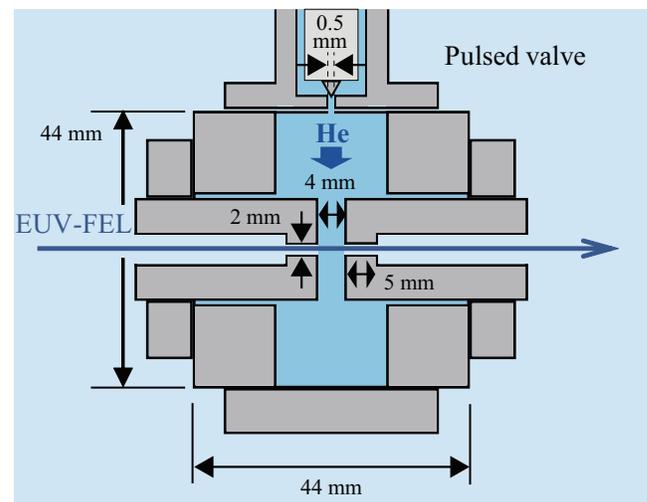}
        \end{center}
	\caption{Cross-section of the helium gas target. 
        }
	\label{fig:target}
\end{figure}
%+++++++++++++++++++++ Figures 2&3  +++++++++++++++++++++%

%------------------------------------- Subsection -------------------------------------%
\subsection{Beam line}
%------------------------------------- Subsection -------------------------------------%
The experiment was performed at the beam line of the SPring-8 Compact SASE Source (SCSS) test accelerator facility.
It provides extreme ultraviolet (EUV) radiation produced by a free electron laser (FEL) based on self-amplified spontaneous emission (SASE)\cite{FEL-1, FEL-2}. 
In this experiment, the FEL was tuned to  $\lambda=53.7$ nm with an average energy of $\sim3$ $\mu$J and a pulse duration of $\sim30$ fs.
The spread of the FEL wavelength was about $\sim0.5$ nm.
There were pulse-to-pulse fluctuations both in intensity (typically 10\%) and wavelength profile, which due to the SASE process\cite{FEL-1}  consisted of several sharp spikes with widths less than 0.05 nm.
The FEL beam was focused to give a spot size of 10 $\mu$m at the target by a pair of elliptical and cylindrical mirrors.\cite{FEL-3} 
The main characteristics of the FEL pulses are summarized in Table \ref{Table:EUV-FEL}.

Several beam line elements\cite{FEL-3} were used during the experiments. 
These include the argon gas attenuator (which allows the FEL intensity to be varied between 100 \% and 1 \% by controlling the pressure of argon gas, and was used in the timing calibration with a fast photodiode to synchronize the fs-laser pulse to the FEL, see Fig.\ \ref{fig:Results-raw}.), and the argon ionization chamber, which was used to monitor the total FEL intensity on a pulse-by-pulse basis. 
Since argon has a continuous ionization cross-section in the EUV region, the collected ion current is directly proportional to pulse energy. 
A new beamline component was developed specifically for these experiments (see Appendix 1 for a detailed description). 
This helium resonance monitor can be used to determine pulse-by-pulse the proportion of the FEL pulse energy which falls within the transition lineshapes of the helium excited levels. 
Finally, a femotosecond (fs) optical parametric amplifier pumped by a Ti:Sapphire laser\cite{FEL-3} was synchronized to the FEL pulses to provide a time calibration signal for the photodetectors. 
This allowed us to accurately determine the SR delay time.

%+++++++++++++++++++++ Table I : FEL characteristics +++++++++++++++++++++%
\begin{table}[h]
\caption{SCSS EUV-FEL characteristics.}
\label{Table:EUV-FEL}
\begin{center}
\begin{tabular}{r|r|l}   \hline \hline
  Parameter &  Value &  Remarks  \\ \hline
  Wavelength of radiation & 53.7 nm &  51--61 nm available\\ 
  Spectral width & $\sim$1 \%  &   \\
  Pulse energy & $\sim$3 $\mu$J & $\sim$30 $\mu$J maximum  \\
  Pulse width (FWHM) & $\sim$ 30 fs & \\
  Intensity fluctuation (typical) & 10\% & \\
  Repetition rate & 30 Hz & \\
    \hline \hline
 \end{tabular}
\end{center}
\end{table}
%+++++++++++++++++++++ Table I +++++++++++++++++++++%

%------------------------------------- Subsection -------------------------------------%
\subsection{Helium gas target}
%------------------------------------- Subsection -------------------------------------%

Figure \ref{fig:target} shows a cross sectional view of the target cell located at the FEL focus point in the vacuum chamber. 
It consists of a pulsed solenoid gas valve (General Valve Series 9, Parker Haniffin, with a 0.5-mm-diameter straight channel) synchronized to the FEL pulses, entrance and exit apertures of 2-mm-diameter, and auxiliary space for pumping. 
The target length (4 mm) is defined by the distance between the two apertures.
In order to vary the helium gas density, the nozzle stagnation pressure, the delay from the FEL trigger signal, or the valve opening duration may be varied.
Whereas the delay and duration were varied to obtain calibration data, the data reported here was recorded with these parameters fixed. 
To record data at different gas densities the nozzle stagnation pressure was varied between $4.7 \times 10^{2}$ Pa and $1.8 \times 10^{5}$ Pa.

The gas density in the cell, $N_0$, was estimated by solving gas flow equations incorporating parameters such as the target geometry, conductance, and pumping speed.
The calculation results were compared to the actual pressure data monitored at several  
points outside the target cell.
Since the time responses of the pressure gauges were slow compared to the valve repetition rate of 30 Hz, we conservatively assign uncertainties on the calculated gas densities used below which are 3 times larger than those estimated from the flow calculations, although the agreement between measured values and the simulation results is good.
This factor of 3 was estimated by comparison with additional reference measurements made using argon gas (instead of helium), and measuring the absorbance of the FEL pulses as a function of delay from the trigger.

%------------------------------------- Subsection -------------------------------------%
\subsection{Detectors}
%------------------------------------- Subsection -------------------------------------%
The detector system was located downstream of the gas cell along the FEL beam line path.
The SR signals were monitored through an optical window outside the chamber, as shown in Fig.\ \ref{fig:Setup}.
We used three identical fast photodiodes (PD) (G4176-03, Hamamatsu Photonics K.K.) to detect simultaneously the radiation from the excited levels in the visible wavelength region.
The cathode area and response time of the PDs were 0.04 mm$^2$ and 30 ps, respectively.
Beamsplitters (BSW10 and BSS10, Thorlabs) and Ag mirrors (PF10-03-P01, Thorlabs) were used to distribute SR pulses of different wavelengths to three separate optical paths.
In front of each PD, we placed a band-pass filter (FB500-10, FB670-10, or FB730-10, Thorlabs) with a peak transmission wavelength corresponding to one of the three SR transitions.
The separated SR pulses were collimated by $f$=25 mm lenses (LB1761A, Thorlabs) into each detector.
Bias voltages were supplied through bias-tees (5550B-104, Picosecond Pulse Labs).
The quantum efficiencies including losses due to the optics mentioned above were estimated to be
0.104, 0.111, and 0.114 e/photon for the 502, 668, and 728-nm detectors, respectively.
An additional detector of the same model was used for timing calibration by the synchronised fs-laser pulses.
The procedure used to determine the time origin was as follows. 
First, the delay between the fs-laser pulse and the FEL pulse at the sample position was set to zero using a separate measurement. 
The synchronized fs-laser pulse was then split into two paths before entering the beamline, with one directed to the additional fast photodiode and the other merged with the FEL path through the gas cell. 
By removing the bandpass filters and detecting the fs-laser pulses on each PD, the time origin of each detector could be determined by comparison with the directly detected fs-laser pulse. 
The overall time calibration accuracy was limited by the time resolution of the detection system to a few picoseconds.

%------------------------------------- Subsection -------------------------------------%
\subsection{Data acquisition system}
%------------------------------------- Subsection -------------------------------------%
The signals from the photodetectors were recorded using an oscilloscope with a bandwidth of 12 GHz (DS091204A, Agilent Technologies, Inc.), and transferred to a computer for later analysis, together with auxiliary data such as the signals from the argon and helium monitors.
25000 pulses were accumulated at each parameter setting.
We took data at 7 different gas densities $N_0$: $6 \times 10^{14}$, $2 \times 10^{15}$, $5 \times 10^{15}$, $1 \times 10^{16}$, $4 \times 10^{16}$, $3 \times 10^{17}$, and $6 \times 10^{17}$ cm$^{-3}$.

%------------------------------------- Section 3: Experimental Results -------------------------------------%
\section{Experimental Results} 
%------------------------------------- Section 3 -------------------------------------%

%------------------------------------- Subsection -------------------------------------%
\subsection{Event selection and averaging}
%------------------------------------- Subsection -------------------------------------%

%+++++++++++++++++++++ Figure 4 : Raw data +++++++++++++++++++++%
\begin{figure}[h]
	\begin{center}
		\includegraphics[width=8.5cm]{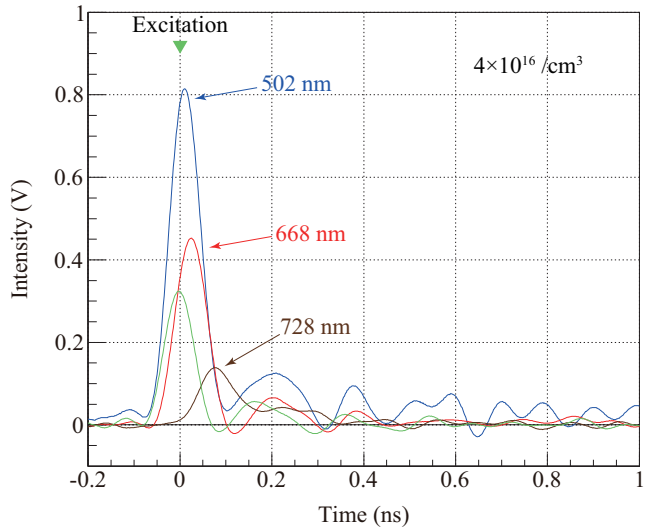}
        \end{center}
	\caption{
	Typical super-radiance pulses recorded in a single FEL shot for all three wavelengths. 
	The time axis was calibrated by the fs-laser pulse (shown as a green curve) which was synchronized to the FEL.
	}
	\label{fig:Results-raw}
\end{figure}
%+++++++++++++++++++++ Figure 4 +++++++++++++++++++++%

A typical example of the traces recorded from a single FEL pulse is shown in Fig. \ref{fig:Results-raw}. 
All three SR transitions are observed. 
The helium density was $4 \times 10^{16}$ cm$^{-3}$, and the ringing which can be seen is due to the detection electronics.

%+++++++++++++++++++++ Figure 5 : Results, all +++++++++++++++++++++%
\begin{figure}[h]
	\begin{center}
		\includegraphics[width=8.5cm]{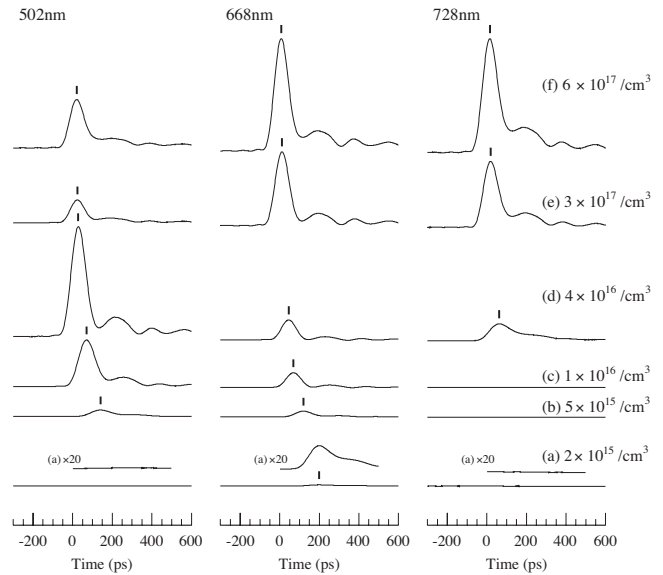}
        \end{center}
	\caption{
	Averaged time profiles at each helium density (a) to (f).
	Left, 502 nm; center, 668 nm; right, 728 nm.
	The vertical scales are normalized to the maximum peak intensity observed for each wavelength.
	The sticks indicate the peak positions (delay time) used in plotting Fig.\ \ref{fig:Results-delaytime}.
	}
	\label{fig:Results-all}
\end{figure}
%+++++++++++++++++++++ Figure 5  +++++++++++++++++++++%

Figure \ref{fig:Results-all} shows the detector outputs averaged over all selected events.
Event selection was performed using the resonance helium gas monitor signals, 
since only those events in which the FEL pulses have sufficient resonant energy are expected to lead to super-radiance decay. 
We used a cut condition of $C_{\mathrm{He}} > 30$ per FEL pulse for event selection, where $C_{\mathrm{He}}$ is the number of pulse counts in the resonance monitor signal (see the appendix).
At the helium density of $N_0 = 6 \times 10^{14}$ cm$^{-3}$, we detected weak signals only at 668 nm; thus we omitted the data set in Fig.\ \ref{fig:Results-all}.

%------------------------------------- Subsection -------------------------------------%
\subsection{Delay and pulse area}
%------------------------------------- Subsection -------------------------------------%

%+++++++++++++++++++++ Figure 6 : Results, delay +++++++++++++++++++++%
\begin{figure}[h]
	\begin{center}
		\includegraphics[width=8.5cm]{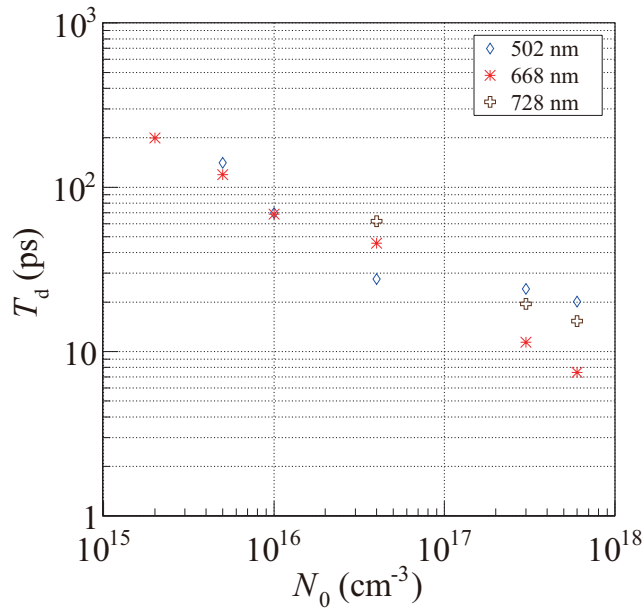}
        \end{center}
	\caption{SR pulse delay time $T_{\mathrm{d}}$ vs helium density in the cell $N_0$ for each SR wavelength.}
	\label{fig:Results-delaytime}
\end{figure}
%+++++++++++++++++++++ Figure 6 +++++++++++++++++++++%

\begin{figure}[h]
	\begin{center}
		\includegraphics[width=8.5cm]{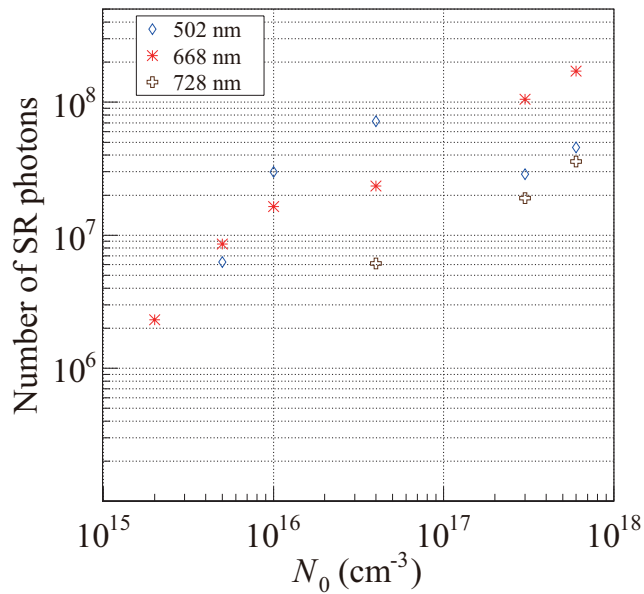}
        \end{center}
	\caption{Number of SR photons vs helium density in the cell $N_0$ for each SR wavelength.}
	\label{fig:Results-pulse-area}
\end{figure}

Figures \ref{fig:Results-delaytime} and \ref{fig:Results-pulse-area} show the average delay times and photon numbers of the observed SR pulses for each of the different helium densities $N_0$.
%\footnote{
($^{1}$ In this work we have not performed a correlation analysis
between super-radiance at different wavelengths for each FEL pulse. To
focus on the sample density dependence, we only consider averaged
data.)
%}
The delay time $T_{\mathrm{d}}$ was defined as the delay of the peak of the pulse relative to the origin, and the photon numbers calculated by integrating the average waveforms and correcting for each detector's quantum efficiency.
No data is plotted for the 502 nm pulses at $N_0 \leq 2 \times 10^{15}$ cm$^{-3}$, or the 728 nm pulses at $N_0 \leq 1 \times 10^{16}$ cm$^{-3}$, since no SR peaks could be distinguished from the background level at these densities.
The statistical errors associated with the delay time and power measurements are less than 5\% and 10\%, respectively.

%+++++++++++++++++++++ Figure 7 : Results, pulse area +++++++++++++++++++++%
%+++++++++++++++++++++ Figure 7 +++++++++++++++++++++%

%------------------------------------- Section 4: Discussions and Summary -------------------------------------%
\section{Discussions and Summary}
%------------------------------------- Section 4 -------------------------------------%
Several features are prominent in the obtained data of Figs.\ \ref{fig:Results-all}, \ref{fig:Results-delaytime}, and \ref{fig:Results-pulse-area}.
First of all, only the 668 nm radiation appears at the lowest target density.
Also, in contrast to the 668 nm and 728 nm pulse heights, the 502 nm pulse heights do not increase monotonically with increasing density.
Finally, at the highest densities the 668 nm SR photon numbers are more than five times higher than those of the other two wavelengths.
These features are discussed below.

The simple semiclassical SR theories \cite{SR-review-Benedict, SR-experiment-Skribanowitz} 
for a two-level system predict that the entire SR process can be characterised by a single parameter.
This parameter has units of time and can be written as
\begin{eqnarray}
T_{\mathrm{R}} = \frac{8 \pi}{3 N A L \lambda^2},
\label{eq:SR}
\end{eqnarray}
where $N$ is the excited target density, $A$ and $\lambda$ the spontaneous radiative transition rate (
Einstein's A-coefficient) and wavelength, and $L$ the target length.
The SR time profile is expected to have a peak with a delay time $T_{\mathrm{d}} = T_{\mathrm{R}}/4 | \ln(NV/2 \pi) |^2$ where $V$ is the target volume.

From the $1s3p$ state, three deexcitation paths (that is, to $1s2s$, $1s3s$, and $1s3d$) exist 
besides that to the ground state $1s^2$.
For a given $N$ and $L$, the value of $A \lambda^2$ is a measure of the 'readiness' for SR to occur on a particular transition (see Eq.\ (\ref{eq:SR})).
Transitions with larger values of $A \lambda^2$ can be expected to develop as SR more readily. %\footnote{
($^{2}$All the densities are well above the thresholds for SR defined by the Doppler dephasing in the two-level model:
$T_{\mathrm{R}} \ll T_{\mathrm{Doppler}}$ (the Doppler dephasing time).)
%}
The values of $A \lambda^2$ for the three relevant transitions here are 
$3.4 \times 10^{-6}$ ($1s3p \to 1s2s$), $1.4 \times 10^{-5}$ ($1s3p \to 1s3s$), and $1.4 \times 10^{-6}$ $\mathrm{m}^{2}\,\mathrm{s}^{-1}$ ($1s3p \to 1s3d$) (see Table \ref{Table:transitions}).

Thus, according to the two-level model, SR can be expected to develop first on the $1s3p \to 1s3s$ transition as the sample density is increased above values where no SR occurs.
Following the transfer of population to the $1s3s$ state, it can be expected that a second SR decay develops on the $1s3s \to 1s2p$ transition at 728 nm ($A \lambda^2 = 9.7 \times 10^{-6}$ $\mathrm{m}^{2}\,\mathrm{s}^{-1}$ for this transition).
The $1s3p \to 1s2s$ transition at 502 nm can be expected to be the next SR transition to develop.
The SR decay at a wavelength of 668 nm on the $1s3d \to 1s2s$ transition is expected to be the last to appear since the $1s3p \to 1s3d$ transition would not occur as SR until the sample density becomes high enough.

It is clear that the predictions of this simple model, however, contradicts the experimental results.
The observation of SR only at 668 nm at the lowest densities suggests that either the $1s3p \to 1s3d$ transition is strongly enhanced, or that the $1s3d$ state can be directly populated by the FEL pulses.
This can be expected to occur if coupling exists between the upper states, since the linewidth of the FEL pulses is broader than the separation between the states. 
Further, the short delay times $T_{\mathrm{d}}$ observed for the 668 nm and 728 nm transitions also suggests that the $1s3s$ and $1s3d$ states may be being populated directly, since two-step cascade processes would lead to longer observed delays due to the delay of the (undetected) first step of the cascade.

However, even if we assume that the $1s3d$ and $1s3p$ states can be populated with equal probability to that of the $1s3p$ state, the simple two-level model still fails to explain the observed behaviour. 
Comparing the values of $A \lambda^2$ for just the 668 nm, 502 nm, and 728 nm transitions, this assumption would predict an SR appearance order of 668 nm $\to$ 728 nm $\to$ 502 nm, with increasing sample density. 
The appearance of the 728 nm SR transition only at the highest densities is not explained.
Further, the suppression of the 502 nm SR transition at high densities does not follow from these assumptions.

It is clear that an understanding of the observations requires a treatment beyond the simple two-level model, taking into account couplings between states in the excitation process, and the competitive nature when initial and final states are shared. 
The strong electric field at the focus position can lead to AC Stark mixing of the states. 
In addition, competitive dynamics occurs during the propagation of the SR fields through the gas target. 
To fully understand the experimental data requires numerical simulations of the nonlinear Maxwell-Block equations for a multi-level system.

In summary, we have observed SR transitions at wavelengths of 502 nm, 668 nm, and 728 nm following the excitation of helium atoms by EUV-FEL pulses. 
The order of appearance of the three transitions with increasing target pressure cannot be explained by the  predictions of a simple two-level model based on the characteristic SR time parameter. 
To explain the observed changes in peak intensity among the competitive SR transitions requires an analysis using simulations of the Maxwell-Bloch equations for a multi-level system, which is the subject of an accompanying paper.\cite{Analysis-Paper}

\newpage
%------------------------------------- Acknowledgment -------------------------------------%
\section*{Acknowledgment}
We would like to thank Kentarou Kawaguchi, Itsuo Nakano, Minoru Tanaka, Jian Tang, Satoshi Uetake, Tomonari Wakabayashi, Akihiro Yoshimi, Koji Yoshimura, and Motohiko Yoshimura for valuable discussions during the early stages of the experiment.
We are grateful to the RIKEN SCSS test accelerator engineering team for experimental support and stable operation of the accelerator.
We acknowledge Toshio Horigome for the detector design of the resonance helium gas monitor.
This research was partially supported by a Grant-in-Aid for Scientific Research A (21244032), a Grant-in-Aid for Challenging Exploratory Research (24654132), and a Grant-in-Aid for Scientific Research on Innovative Areas ``Extreme quantum world opened up by atoms'' (21104002) from the Ministry of Education, Culture, Sports, Science, and Technology of Japan.

%------------------------------------- Appendix: helium monitor -------------------------------------%
\section*{Appendix: Resonance helium gas monitor}
%------------------------------------- Appendix: -------------------------------------%

In this Appendix, we describe the details of the resonance monitor\cite{He-SR-Harries} that was used in the experiment to monitor the resonant component in the EUV-FEL pulses.

%------------------------------------- Subsection -------------------------------------%
\subsection*{Basic operation principle}

The resonance monitor developed here utilizes the fact that an atom in 
a
metastable state with energy $E$ can eject one or more electrons from a solid surface when $E$ is larger than the work function.\cite{Hotop1996191}
When a helium atom is excited to the $1s3p$ state, it decays to the $1s2s$ state with a ratio of 2.4 \%.
This lower state is metastable, with an A-coefficient of 51 s$^{-1}$.
Due to the high internal energy ($E \sim 20$ eV), collisions of helium atoms in this state with solid surfaces can lead to the production of one or more free electrons.
When a MCP (microchannel plate) device is used as a solid surface, the electron current is multiplied by subsequent secondary electron emission, giving electronic signals in the anode.
The quantum efficiency of MCP detection is typically around 10 \%.
The excitation probability of helium is proportional to the intensity of the resonant component in the FEL pulse.
Therefore, counting the signals of the resonance monitor gives a measure of the resonant component energy in each pulse.
The SASE-FEL pulses consist of spike-like spectral structures, which have linewidths of $\sim0.05$ nm. 
Thus the resonance monitor is only sensitive to FEL pulses which contain spike components which have peaks at wavelengths within $\pm0.05$ nm of the exact resonance wavelength.

%------------------------------------- Subsection -------------------------------------%

%------------------------------------- Subsection -------------------------------------%
\subsection*{Design}

%+++++++++++++++++++++ Figure A1 : helium monitor design +++++++++++++++++++++%
\begin{figure}[h]
	\begin{center}
		\includegraphics[width=8.5cm]{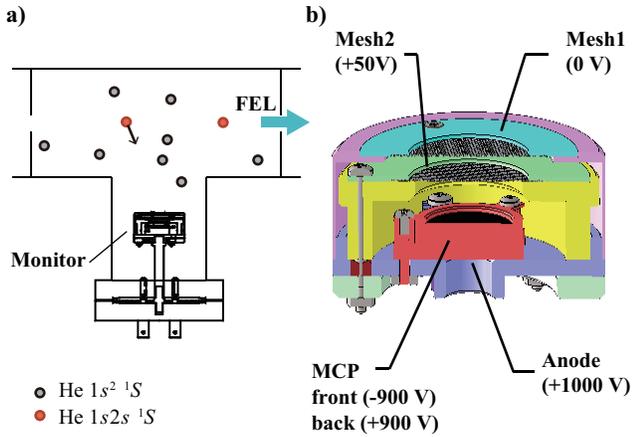}
        \end{center}
	\caption{
	a) Schematic view of the resonance helium gas monitor.
	b) Design of the detection part of the resonance monitor (cross sectional view).
	The numbers in parentheses are the applied voltages to the electrodes.
	}
	\label{fig:He monitor design}
\end{figure}
%+++++++++++++++++++++ Figure A1 +++++++++++++++++++++%

The detector assembly is summarised in Fig. \ref{fig:He monitor design}.
Helium atoms move thermally in the helium monitor chamber.
After they are excited by the FEL, some of them reach the detection assembly (Fig.\ \ref{fig:He monitor design}a).
The second mesh (+50 V) works as a repeller for positive ions (Fig.\ \ref{fig:He monitor design}b).
The front side of the 2-stage MCP (Hamamatsu F4655) was biased by -900 V, working as an electron repeller.
The neutral metastable helium atoms can directly hit the front-side walls of the MCP, and the released electrons are accelerated and multiplied.
The electrons are amplified and accelerated by a 1800 V acceleration voltage, and collected by the anode plate which is biased by + 100 V relative to the end surface of the MCP.
Finally the signals are acquired by a digital oscilloscope (Tektronix DPO7054) after being processed by a RC circuit.

%------------------------------------- Subsection -------------------------------------%

%------------------------------------- Subsection -------------------------------------%
\subsection*{Specifications}

%+++++++++++++++++++++ Figure A2 : helium monitor raw signal +++++++++++++++++++++%
\begin{figure}[h]
	\begin{center}
		\includegraphics[width=8.5cm]{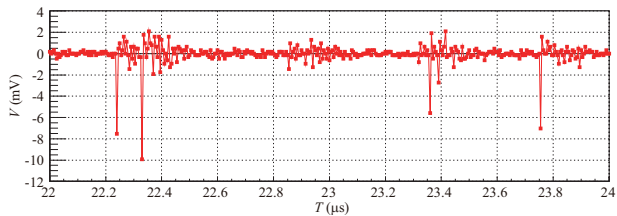}
        \end{center}
	\caption{Resonance monitor raw signals after a FEL pulse at time zero.}
	\label{fig:He monitor raw signal}
\end{figure}
%+++++++++++++++++++++ Figure A2 +++++++++++++++++++++%

Figure \ref{fig:He monitor raw signal} shows an oscilloscope trace of the resonance monitor signal obtained with $P_{\mathrm{He}} = 1.7 \times 10^{-3}$ Pa in the monitor chamber.
Each pulse signal is countable except in the region just after the FEL pulse where the background spike noise from the EUV pulse prevents counting.
Fluctuation of the peak height may arise from the difference of the depth along the MCP where the metastable helium atoms hit the channel surfaces.
Pulses with peak heights $\geq$ 1 mV were counted as the monitor signals.

%+++++++++++++++++++++ Figure A3 : helium monitor pressure dependence +++++++++++++++++++++%
\begin{figure}[h]
	\begin{center}
		\includegraphics[width=8.5cm]{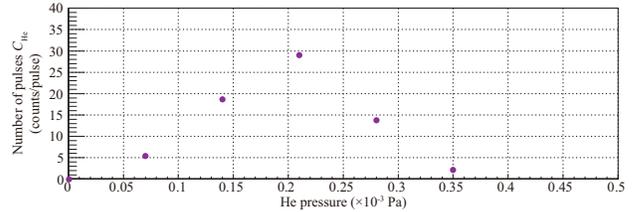}
        \end{center}
	\caption{Helium pressure dependence of the 
	helium resonance monitor signal.
	}
	\label{fig:He monitor pressure dependence}
\end{figure}
%+++++++++++++++++++++ Figure A3 +++++++++++++++++++++%

There is an optimal helium pressure for the resonance monitor.
Figure \ref{fig:He monitor pressure dependence} shows the counting rate ($C_{\mathrm{He}}$) as a function of the helium pressure.
As seen, it shows a broad peak structure which is understood by the following observations.
The probability that the resonant component of the FEL pulse is absorbed is given by the Beer's law.
The detector is located at $\sim 50$ cm from the upstream entrance to the monitor chamber, and only helium atoms excited close to the MCP can be detected.
When the pressure is too high, the FEL pulses lose their resonant intensity before reaching the region where atoms can be detected.
Therefore, the helium pressure should be controlled in the initial linear range of Fig.\ \ref{fig:He monitor pressure dependence}, as actually done in the SR measurement ($P_{\mathrm{He}} = 2.1 \times 10^{-4}$ Pa).

%------------------------------------- Subsection -------------------------------------%
\subsection*{Cut condition for data selection}

%+++++++++++++++++++++ Figure A4 : helium monitor cut condition +++++++++++++++++++++%
\begin{figure}[h]
	\begin{center}
		\includegraphics[width=8.5cm]{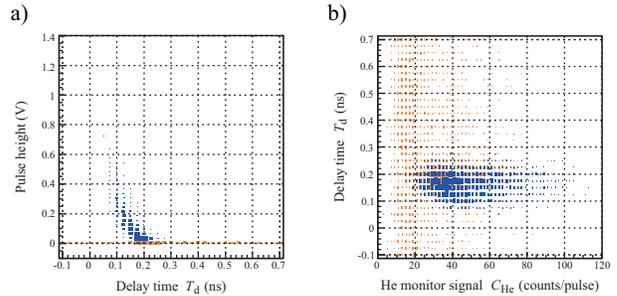}
        \end{center}
	\caption{
Correlation of 502 nm SR and the resonance monitor signals. 
a) Pulse height vs delay time for the 502 nm detector. 
b) 502 nm delay time vs resonance monitor signal counts. 
The blue points show the clear correlation
}
	\label{fig:He monitor cut condition}
\end{figure}
%+++++++++++++++++++++ Figure A4 +++++++++++++++++++++%

Data selection utilised the resonant monitor signals.
Figure \ref{fig:He monitor cut condition}a shows a scatterplot of the pulse peak height vs delay time for a typical 502 nm dataset, constructed from single-shot traces. 
It is clear that two groups of data points exist; data originating in SR (blue points), and background signal (orange). 
Figure \ref{fig:He monitor cut condition}b is a plot for the same dataset, but shows the 502 nm delay time vs the resonant monitor signal. 
For low values of $C_{\mathrm{He}}$ (less than 30, for example), nearly every event is background signal. 
The large correlation between the detection of 502 nm SR and the resonant monitor signal is evidence of the monitor's sensitivity to the resonant component of the FEL pulses. 
A cut condition of $C_{\mathrm{He}} > 30$ was used for the data analysis presented in this paper.

%==============  bibliography ======================================
%---------------------------------------------------%
%                    References
%---------------------------------------------------%
%\bibliographystyle{jpsj}
%\bibliography{HeliumSuperRadianceExperiment_Kuma}

\end{document}